\documentclass[preprint]{aastex}
\usepackage{emulateapj5}
\usepackage{epsfig}

\makeatletter
\newenvironment{inlinefigure}{%
\def\@captype{figure}%
\noindent\begin{minipage}{0.999\linewidth}\begin{center}}
{\end{center}\end{minipage}\smallskip}
\makeatother

\newcommand{\etal}{\mbox{et al.}}
\newcommand{\ksxrb}{\mbox{KS 1731$-$260}}
\newcommand{\aqlxone}{\mbox{Aql X-1}}

\newcommand{\sixb}{\mbox{4U 1636$-$536}}
\newcommand{\sevenb}{\mbox{4U 1702$-$429}}
\newcommand{\slowb}{\mbox{4U 1728$-$34}}

\newcommand{\mxbecl}{\mbox{MXB 1659$-$298}}
\newcommand{\saxms}{\mbox{SAX J1808.4$-$3658}}
\newcommand{\asmms}{\mbox{XTE J0929$-$314}}

\newcommand{\rxte}{{\it RXTE}}

\newcommand{\msun}{\mbox{$M_{\odot}$}}
\newcommand{\rms}{rms}

\newcommand{\program}[1]{{\tt {#1}}}
\newcommand{\html}[1]{{\tt http://#1}}

\shortauthors{Muno, \"{O}zel, \& Chakrabarty}
\shorttitle{Energy Dependence of Burst Oscillations}

\slugcomment{Submitted to \apj.}

\begin{document}
\title{The Energy Dependence of Millisecond Oscillations in Thermonuclear X-ray Bursts}
\author{Michael P. Muno,\altaffilmark{1} Feryal
\"{O}zel,\altaffilmark{2} and Deepto Chakrabarty\altaffilmark{1,3}}

\altaffiltext{1}{Department of Physics and Center for Space Research,
       Massachusetts Institute of Technology, Cambridge, MA 02139,
       muno@space.mit.edu, deepto@space.mit.edu}
\altaffiltext{2}{Hubble Fellow; Institute for Advanced Study, Einstein Dr.,
       Princeton, NJ 08540}
\altaffiltext{3}{Alfred P. Sloan Research Fellow}

\begin{abstract}
We examine the energy-resolved pulse profiles of 51 flux oscillations
observed during the decline of thermonuclear X-ray bursts from 
accreting weakly-magnetized
neutron stars with the {\it Rossi X-ray Timing Explorer}. We find that the
fractional rms amplitudes of the oscillations increase as a function of
energy by 0.25\% keV$^{-1}$ to 0.9\% keV$^{-1}$ between
$3-20$~keV, and are as large as 20\% in the $13-20$~keV band. We also show
that the pulses observed in the higher energy bands generally lag
behind those in lower energy bands by $0.002$ cycles ${\rm
keV}^{-1}$ to $0.007$ cycles ${\rm keV}^{-1}$ between $3-20$~keV. 
This amounts to total delays of
0.03--0.12~cycles between the lowest and highest energy bands, or 
time delays that range from 100--200 $\mu$s. We then model the
oscillations as flux variations arising from temperature patterns on the
surfaces of rapidly rotating neutron stars. In this framework, we find
that the increase in the pulse amplitude with photon energy can be
explained if the cooler regions on the neutron star emit in
the lower energy bands, reducing the flux variations there. On the 
other hand, the Doppler shifts caused by the rapid rotation 
of the neutron star should cause the hard pulses to precede the soft pulses 
by about $0.05$~cycles (100 $\mu$s),
in contrast to the observations. This suggests that the photons
originating from the stellar surface are reprocessed by a hot corona of
electrons before they reach the observer.
\end{abstract}

\section{Introduction}
Thermonuclear bursts from accreting weakly-magnetized neutron stars
provide an excellent opportunity to study emission originating from the 
stellar surface (see Lewin, van Paradijs, \& Taam 1993 for a review). 
For instance, the unstable helium burning that produces the bursts may leave 
asymmetries in the surface brightness of the star that can be observed as 
pulsations at the stellar spin frequency \citep[e.g.,][]{fw82,str97}. 
Indeed, oscillations that are apparently produced by bright regions on 
rapidly rotating neutron stars
now have been observed during bursts from ten neutron star 
low-mass X-ray binaries (LMXBs) with the {\it Rossi X-ray Timing Explorer} 
\citep[\rxte; see][for a review]{str01}. Their frequencies
range between 270--620~Hz \citep{mun02}, which are consistent with the 
assumption that these LMXBs are the progenitors of recycled millisecond 
radio pulsars \citep{alp82,rs82}. Most importantly, after a small frequency 
drift is accounted for, the oscillations are extremely coherent. The 
oscillations have quality factors $Q = \nu/\Delta\nu > 1000$ 
during individual bursts (Strohmayer \etal\ 1996; Strohmayer \& Markwardt
1999, 2002; but see Muno \etal\ 2002a)\nocite{str96,sm99,sm02,mun02}, 
and they appear at frequencies that are stable to a few parts in 1000 in 
bursts observed over several years from any given source 
\citep{str98b,gil02,mun02}. 
%In one instance, oscillations remained coherent 
%for 600~s during an hour-long ``super-burst'', and exhibited a slight
%frequency drift that was consistent with Doppler shifts due to the 
%known orbit of the binary system \citep{sm02}. 
The stability and coherence
of these oscillations provides compelling evidence that they originate from
brightness patterns on rotating neutron stars.

Burst oscillations therefore allow one to study a range of questions, 
including the physics of nuclear burning
on the stellar surface, the effects of gravitational self-lensing by the
neutron star, and the equation of state of dense nuclear matter. The most
constraining data so far have come from the amplitudes and the profiles of the 
oscillations. The fractional \rms\ amplitudes of the oscillations can be as 
large as 50\% during the first second of a burst \citep{str98a}. 
If the oscillations result 
from two
antipodal bright regions on the surface, as has been proposed based upon 
both observational and theoretical considerations (Miller, Lamb, \& Psaltis
1998; Miller 1999)\nocite{mlp98,mil99}, then
amplitudes this large can only be produced if the neutron star is larger than
$R/R_{\rm Sch} = 3.1$ (Miller \& Lamb 1998; Nath, Strohmayer, \& Swank 
2002)\nocite{ml98,nss02}, where $R_{\rm Sch}$ is the 
Schwarzschild radius. Moreover, the profiles
of the oscillations are nearly perfectly sinusoidal --- the amplitudes of 
the harmonic components are less than 0.5\% of those of the main signals.
This suggests that the patterns forming on the stellar surfaces are either
highly symmetric or restricted to near the rotational poles 
(Muno, \"{O}zel, \& Chakrabarty 2002b)\nocite{moc02}. 

The surface velocity of the neutron star ($\approx 0.1c$ for a 10 km  
star spinning at 500~Hz) can be measured from variations in the 
amplitude and phase of the oscillations as a function of energy, which
would provide another constraint on the stellar radius. The amplitudes of the 
oscillations should increase as a function of energy because Doppler 
shifts cause the Wien tail of the spectral distribution to move in and 
out of the highest energy bands \citep{ml98}. However, the amplitudes will 
also be set by the surface temperature distribution, since 
a finite surface temperature produces a constant background against 
which a hot region is less distinct at low energies \citep{page95}. 
The relative phases of the waveforms as a function of energy, on the other
\begin{figure*}[thb]
%\epsscale{0.8}
%\plotone{f1.eps}
\centerline{\epsfig{file=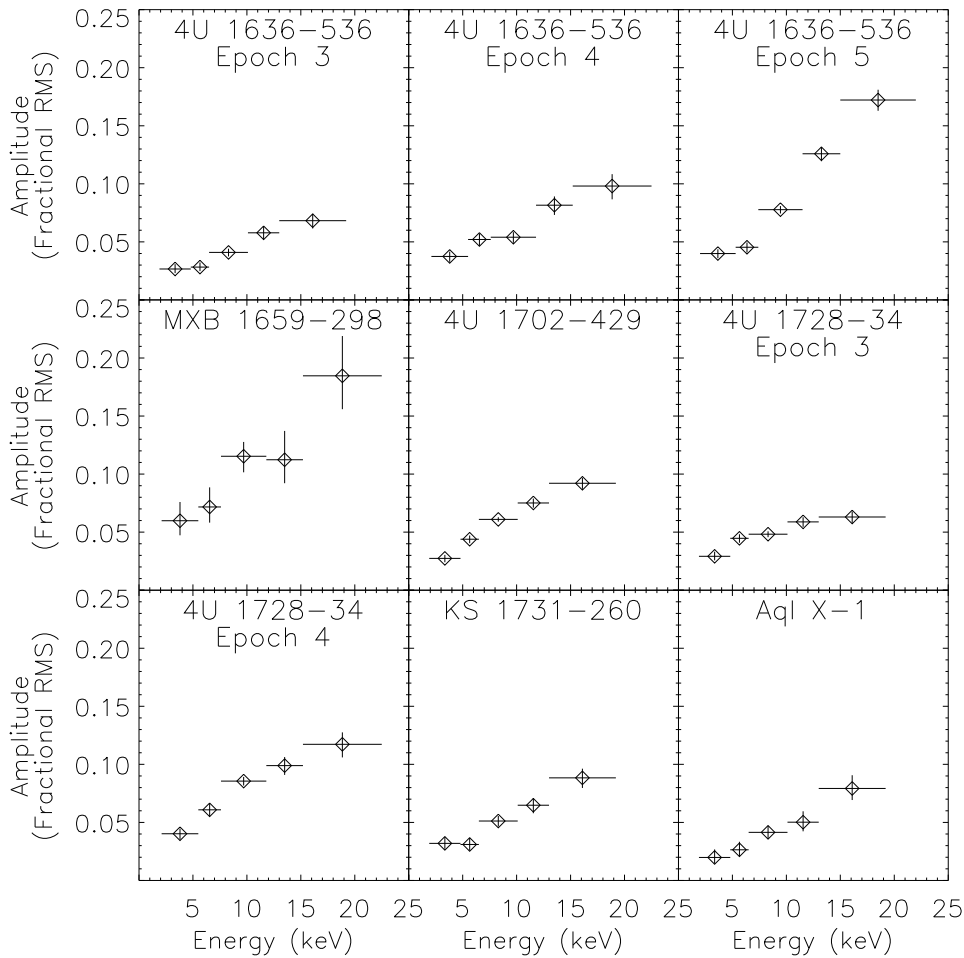,width=0.725\linewidth}}
\caption{
The energy dependence of the amplitudes of the averaged oscillation 
profiles from six sources. For each source, oscillations were only 
averaged for a single epoch during which the gain of the PCA was
relatively constant. In all cases, the amplitudes increase significantly
as a function of the photon energy.}
\label{fig:ampve}
\end{figure*}
hand, are affected only by the Doppler shifts of the tail of the 
spectrum, which move the peak of the high-energy pulse to an earlier phase 
(Braje, Romani, \& Rauche 2000; note that Doppler beaming
and photon arrival time delays also shift the overall phase of the entire
pulse)\nocite{brr00}. Evidence that the hard pulse leads 
the soft has been reported 
in a burst oscillation from \aqlxone\ by \citet{for99}, although further 
studies question whether this is due to rotational motion \citep{fox00}. 

In this paper, we present a study 
of the amplitudes and phases of burst oscillations as a function of 
photon energy. In Section~\ref{sec:obs} we analyze the energy-resolved
waveforms of 59 oscillations observed with the Proportional Counter Array 
(PCA) on \rxte. In Section~\ref{sec:mod}, we present theoretical waveforms 
produced from the models of \citet{moc02}, taking into account the response
of the PCA to facilitate comparison with the data. We consider several simple 
temperature distributions on the neutron star. In Section~\ref{sec:comp},
we explore whether we can measure Doppler effects in the observed 
waveforms. 

\section{Observed Waveforms \label{sec:obs}}

As part of an ongoing study of thermonuclear bursts observed
with the \rxte\ Proportional Counter Array \citep[PCA;][]{jah96}, 
we identified 68 X-ray bursts containing millisecond oscillations 
in data from eight sources that were in the public domain as of 
September 2001 \citep{mun02}. The PCA is composed of 
five gas-filled proportional counter units (PCUs) that are sensitive to X-rays 
between 2.5--60~keV with an energy resolution of about 15\%. Between 
2 and 5 PCUs were active during the observations in our sample. Each 
PCU has an effective area of 1200 cm$^2$ and is capable of recording 
photon arrival times with microsecond time resolution. For most of the 
observations, data with both $2^{-13}$~s (122 $\mu$s) time resolution and 
between 8--64 energy channels were recorded on an event-by-event basis, and 
were used to study the profiles of the oscillations.

We first modeled the frequency evolution of the oscillations observed in the 
entire PCA bandpass using a phase connection technique similar to that 
used in pulsar timing \citep{mt77,mun02}. To implement this technique, 
we folded the data in short intervals (0.25--0.5 s) using a 
trial phase model, which was then refined through a least $\chi^2$ fit to 
the 
residuals. This provides excellent frequency resolution and a statistical 
measure of how well the model reproduces the data. We have previously used the 
resulting phase models to study the frequency evolution, amplitudes, and 
profiles of the burst oscillations \citep{mun02,moc02}.

We then applied the best-fit phase models to produce folded profiles in 
7 energy channels. Since the gains of the proportional counters have been 
changed several 
times during the mission, the energies corresponding to the detector channels 
vary. Data with sufficient energy and time resolution were only
used to record burst oscillations in gain epochs 3 (1996 April 15 through
1999 March 22), 4 (1999 March 22 through 2000 May 13), and 5 
(2000 May 13 to present).
We were unable to choose detector channels with identical energy 
boundaries in different gain epochs, since the high time resolution data were 
usually taken with too few energy channels. Therefore, we defined the energy 
boundaries by detector channels 5, 13, 18, 28, 36, and 53. These boundaries 
roughly correspond to $<$2, 5, 7, 10, 13, and $>$20~keV during gain epoch 
3 (see also Table~\ref{tab:ebounds}). 
The low ($< 2$~keV) and high
($> 20$~keV) energy channels generally were dominated by background events, 
and were therefore ignored during our analysis. We found that 51 
oscillation trains from 6 sources were sampled with adequate time and 
energy resolution for this study. The sources, oscillation frequencies, and
numbers of oscillations examined are listed in Table~\ref{tab:sum}. 

We note that the energy-resolved data that we used (Table~\ref{tab:sum}) 
contained up to 30\% fewer counts than the profiles created for a 
single energy channel in \citet{moc02}. This is because the \rxte\ data 
buffers could only record energy-resolved data for a fraction of each 
second before they were filled during the brightest segments of many bursts.
Other modes designed specifically to record bursts were available for producing
the frequency models (so-called burst-catcher modes), but sacrificed energy 
resolution in favor of high time resolution. The burst-catcher modes were not 
used to produce energy-resolved waveforms.

We measured the amplitudes of the oscillations as a function of energy by 
computing a Fourier power spectrum of the folded profiles in units of 
total counts per phase bin. We normalized 
the power according to \citet{lea83}, so that the fractional \rms\ amplitude 
at any multiple of the oscillation frequency is
\begin{equation}
A_n = 
\left({{P_n}\over{I_\gamma}} \right)^{1/2}{{I_\gamma}\over{I_\gamma-B_\gamma}},
\label{eq:acont}
\end{equation}
where $P_n$ is the power at the $n$th bin of the Fourier spectrum, 
$I_{\gamma}$ is the total number of counts in the profile, and $B_\gamma$
is the estimated number of background counts in the profile.
The background was taken to be the persistent emission during the 15 s 
prior to the burst.
This equation is valid so long as the phase and frequency of the 
oscillation is known, as it is by design for our folded profiles.
Uncertainties and upper limits on the amplitudes are computed taking into
account the distribution of powers from Poisson noise in the spectrum, 
using the algorithm in the Appendix of \citet{vau94}.
We measured the phases of the oscillations by linear least-squares fits of
sinusoids to the folded profiles. The uncertainties on the phases 
were determined from the diagonal values of the covariance matrices that were
derived as part of the least-squares algorithm \citep{pre92}. 

We also summed the energy-resolved profiles for all of the oscillations 
that occurred during an individual gain epoch from each source 
(Table~\ref{tab:sum}).
Although the gain of the PCA drifted even during a single epoch, we
confirmed that the drift was not large enough to affect the summed 
profiles. We did so by folding
a set of simulated profiles (see Section~\ref{sec:mod}) through different 
responses generated by the 
FTOOL\footnote{\html{heasarc.gsfc.nasa.gov/docs/software/ftools/}} 
\program{pcarsp} that spanned gain epoch 3,
measuring the amplitude and phase of each individual oscillation, and 
comparing it to those of the summed profile. The values were identical 
to within less than 1\%.

\subsection{Amplitudes as a Function of Energy}

First, in order to determine whether the oscillation amplitudes in individual 
bursts change
significantly as a function of photon energy, we calculated the
$\chi^2$ statistic under the assumption that the amplitudes are constant:
\begin{equation}
\chi^2 = \sum {\left({A_i - \bar{A}_i}\over{\sigma_{A_i}}\right)^2},
\end{equation}
where $A_i$ is the fractional \rms\ amplitude in each energy band, 
$\bar{A_i}$ is the mean amplitude, and $\sigma_{A_i}$ is the 
uncertainty derived according to the method of 
\citet{vau94}. We find that 34 of 51 oscillations are inconsistent with
a constant amplitude as a function of energy at the 90\% confidence level. 
The remaining oscillations that appear to have constant amplitudes typically 
have fewer counts in the folded profile: the mean number of counts in a 
folded profile is $5.5\times10^4$, while those with constant amplitudes 
have on average $3.5\times10^4$ counts.
 
To quantify how the amplitude varies with energy in these oscillations, 
we fit them with a linear function using a least-squares technique. 
A linear fit is only a crude approximation to the functional form expected
from a hot region on a rotating neutron star (see Section~\ref{sec:mod}),
but it is adequate for making simple quantitative comparisons between 
our observational and theoretical results, given the limited statistics in 
most of our measurements. We found that 28 oscillations were consistent with 
a linear increase
in amplitude as a function of energy. The average of the best-fit slopes for 
these oscillations was 0.8\% keV$^{-1}$. The smallest total
detectable increase in amplitude was about 2\% between the 2~keV and 20~keV
bands, while the increase was as large as 30\% in several oscillations 
from \sixb\ and \sevenb. Only 6 oscillations were inconsistent with both
constant amplitudes and linear trends. Two of these 
(occurring on 1996 December 21 17:30:34 and 1999 June 21 19:13:02 
from \sixb, where these and all subsequent times are in Barycentric
Dynamical Time) exhibit random 
variations in amplitude, while the remaining four 
(occurring on 2000 June 15 05:13:02 and 2000 August 9 09:00:34 
from \sixb, and on 1997 September 20 10:08:27 and 1999 August 19 12:13:22 
from \slowb) %have high signal-to-noise, 
show deviations from a linear trend that are consistent
\begin{figure*}[thb]
%\epsscale{0.8}
%\plotone{f2.eps}
\centerline{\epsfig{file=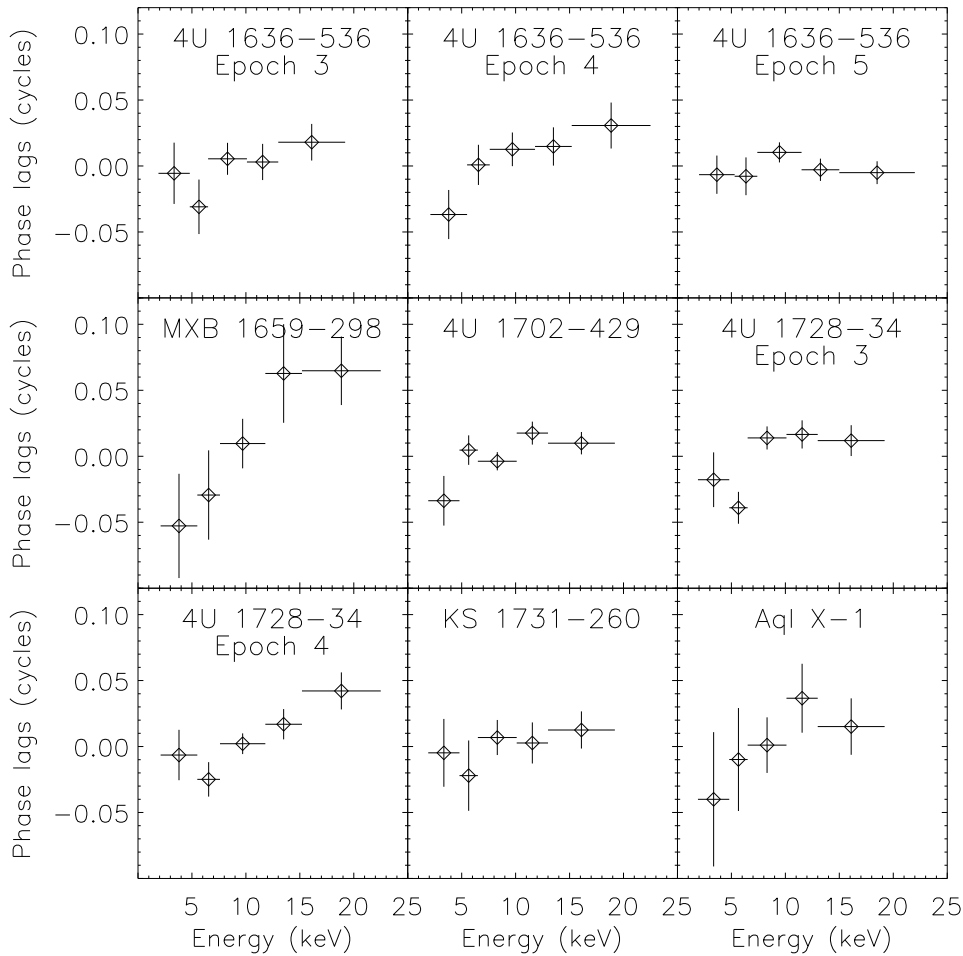,width=0.725\linewidth}}
\caption{
The energy dependence of the phases of the averaged oscillation 
profiles from six sources (compare Figure~\ref{fig:ampve}). Phase zero 
is defined by the profile in the total PCA bandpass. A positive value 
of the phase lag indicates that the profile in that energy range arrives
later than the average profile.}
\label{fig:phave}
\end{figure*}
 with the 
functional form expected from a hot region on a rotating neutron star
(see Section~\ref{sec:mod}).

We then performed the same analysis for 9 summed pulse profiles from 6 
sources.
We list the results in Table~\ref{tab:ampve}, and we display the amplitude as 
a function of energy in Figure~\ref{fig:ampve}. 
The amplitudes from all of the summed profiles increase as a function of 
energy with $> 99.9$\% confidence. In most cases, the increase is consistent
with linear functions with slopes between 0.5--0.8\% keV$^{-1}$. 
This translates to a total increase in amplitude from 5--16\% between 
2--20~keV. Only the 
oscillations from \sixb\ in gain epoch 5 are inconsistent with a linear
trend. These have a higher amplitude in the 2--5~keV band than would be 
expected from extrapolating the trend at higher energies 
(Fig.~\ref{fig:ampve}), but are still roughly consistent with the 
functional form expected from a hot spot.

\subsection{Phase as a Function of Energy}

We then examined the relative phases of the oscillations as a function of
energy. The reference phases were taken to be those of the 
oscillations in the total 2--60~keV bandpass. An increasing value
of the phase lag indicates that the pulse seen at high energies arrives later
than (lags) that at low energies. Of the individual oscillation trains, 
we find 
that only 13 exhibit phases that vary as a function of energy at the 90\% 
confidence level. Of these 13 oscillations, 6 appear to change in phase with 
a slope of $-0.005(3)$ to $0.019(5)$ cycles keV$^{-1}$
(1999 April 29 01:49:24 from \sixb; 1999 April 10 09:54:12 from
\mxbecl; 1999 February 22 04:54:01 from \sevenb; 1999 August 19 15:50:57
from \slowb; and 1996 July 14 04:23:45 and 1999 February 26 17:10:56 from
\ksxrb). Seven 
oscillations are not well-fit by linear trends, but exhibit random 
variations in phase as a function of energy (2000 August 9 01:22:23
and 09:00:34 from \sixb; 1997 July 26 09:09:11 from \sevenb;
1997 September 21 18:10:31, 1997 September 27 11:17:45, 1999 January 31 
21:56:41, and 1999 August 19 12:13:22 from \slowb).
The weighted average 
of the phase lag for the 44 oscillations that can be adequately fit with 
a linear trend is $0.0024(5)$ cycles keV$^{-1}$. 

In Figure~\ref{fig:phave} we plot the phase as a function of energy for
9 summed profiles from 6 sources, and in 
Table~\ref{tab:phave} we compile the results. Applying the $\chi^2$
statistic to the data, we find that the phase varies as a function
of energy at the 90\% confidence level in 5 of the 9 oscillations.
In 4 of these 5 oscillations, the phases are consistent with 
a linear increase as a function of energy with a slope of between 0.002(1) 
and 0.007(3) cycles keV$^{-1}$. Thus, the pulses at 20~keV lag 
behind those at 2~keV by 0.04-0.14 cycles, which translates to time 
delays of 100~$\mu$s (\sevenb) to 200~$\mu$s (\mxbecl). The summed 
oscillation from 
\slowb\ in gain epoch 3 appears to exhibit random variations in 
phase as a function of energy, as is evident in Figure~\ref{fig:phave}.

We note that our sample includes the oscillation from Aql~X-1
in which \citet{for99} reported that the pulse at high energies
{\it precedes} that at low energies in the last half of the oscillation
train. We find that these soft lags are significant only in the interval
that was searched, while earlier the oscillation exhibited hard lags
\citep[see also][]{fox00}.
Over the entire burst, the soft and hard lags cancelled, yielding no net 
delay between
hard and soft photons. We searched for similar changes in the energy
dependence of the phases of other oscillations by dividing the oscillations
into two intervals containing an equal number of counts. We did not find any 
other significant examples of phase lags switching from positive to 
negative.

\section{Model Profiles\label{sec:mod}}

In this section, we examine how the rotation of the neutron star affects the 
energy dependence of the burst oscillations. To that end, we produce 
theoretical 
energy-resolved light curves from circular hot spots on a rotating neutron 
star. We use the techniques outlined by Pechenick, Ftaclas, \& Cohen 
(1983), \citet{brr00}, and Weinberg, Miller, \& Lamb (2001)\nocite{wml01}. 
The amplitudes and profiles of 
our theoretical oscillations in the entire PCA bandpass (2--60~keV) are 
discussed extensively in \citet{moc02}. 

The energy spectra observed during bursts can adequately be modeled as 
blackbody emission with a temperature $kT_\infty$. Therefore, we
use the same spectral distribution to describe the emission
from the neutron star's surface. The angular dependence of the emission 
is described by a Hopf function \citep{cha60}, which is appropriate
for the scattering-dominated atmosphere present during a burst 
\citep[e.g.,][]{mad91}. Photons are propagated to the observer 
through a Schwarzschild metric about a compact object with compactness
$p \equiv R/2M$. 
Corrections to the metric due to the unknown density structure of the neutron 
star should be of order a few percent \citep[e.g.,][]{brr00}, which is of the 
same magnitude as the measurement
uncertainties in the observed waveforms. We assume that the neutron star 
has a mass of 1.4 \msun, and denote its rotation frequency with $\Omega$.

We consider one single and two antipodal circular bright regions with a 
constant temperature, with the rest of the surface taken to emit at a lower 
temperature. 
%We also considered dipolar ($\cos(\theta-\theta_o)$) 
%and quadrapolar ($\cos^2(\theta-\theta_o)$) distributions, where 
%$\theta_o$ was the angle between the peak of the temperature and the 
%observer's line of sight. 
The circular hot region(s) are taken to have angular radius $\rho$ and to be 
centered at an angle $\alpha$ from 
the rotational axis of the neutron star. The angle between the line-of-sight 
of the observer and the spin axis of the neutron star is denoted

\begin{inlinefigure}
%\epsscale{0.5}
%\plotone{f3.eps}
\centerline{\epsfig{file=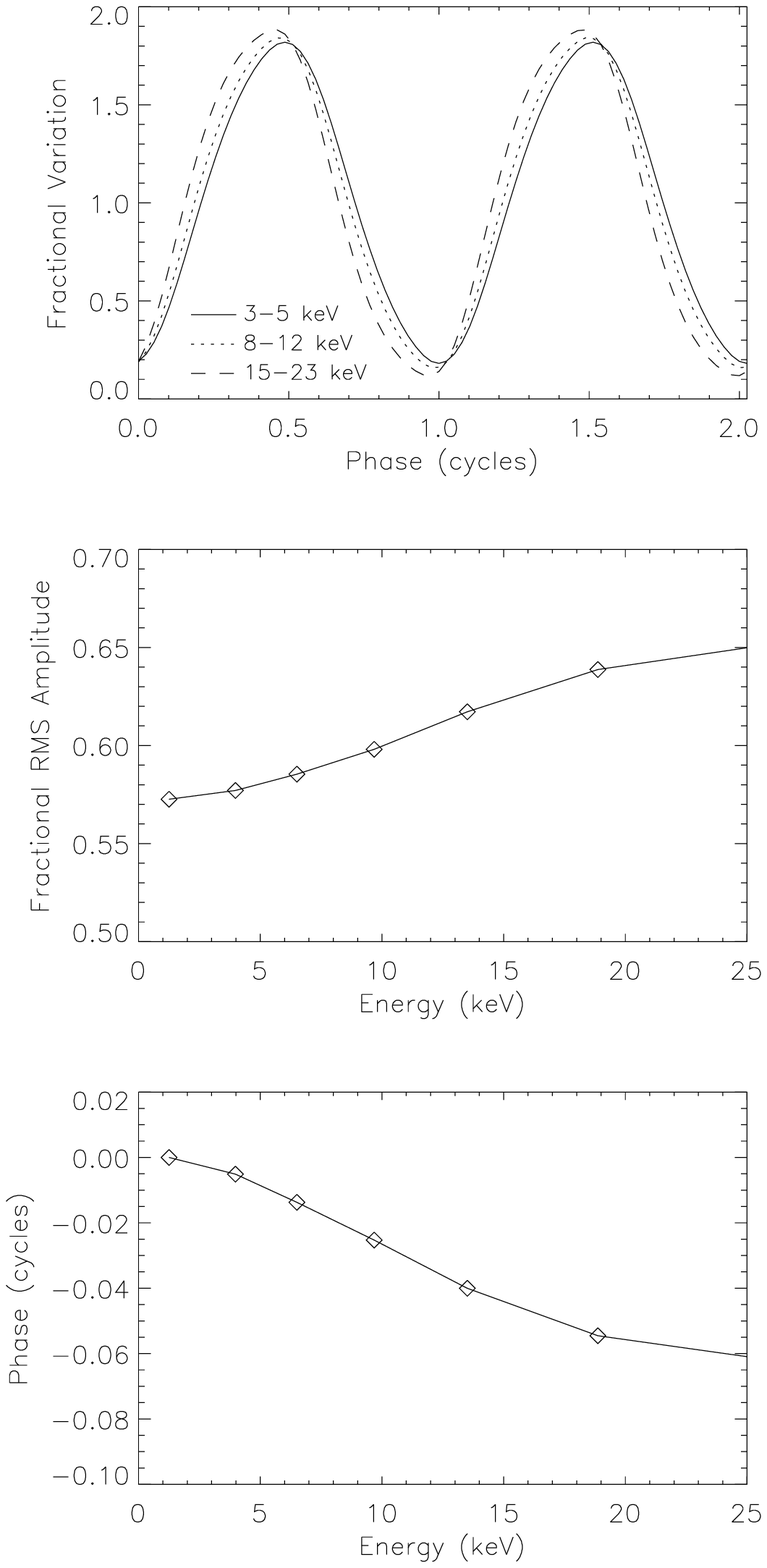,width=0.8\linewidth}}
\caption{Illustration of the energy dependence of the oscillation
profiles from our theoretical models of a bright region on a 
rotating neutron star. We model a 1.4 \msun\ neutron star with a 
radius of 10~km ($p=2.5$) spinning at $\Omega = 600$~Hz. The bright
region has a size $\rho=90^\circ$, is located at $\alpha=90^\circ$, and
is viewed from $\beta=90^\circ$. The response of the \rxte\ PCA was 
accounted for in computing the model profiles. The Doppler and time delay 
effects 
introduced by the rotation of the star cause the amplitude of the oscillation 
increases by 6\% between 1--20~keV, while the phase of the oscillation 
at 20~keV precedes that at 1~keV by 0.055 cycles.}
\label{fig:edep}
\end{inlinefigure}

\noindent
by $\beta$. 
In total, our simulations include seven parameters that can affect
the energy dependence of the oscillations: 
the compactness and the spin frequency of the neutron star; the number, size,
position, and temperature contrast of the hot regions; and the viewing angle
of the observer.  

For each set of parameters, we produce light curves in 40 phase bins
and in 64 energy bins logarithmically spaced between 0.01--25~keV. The
observed spectrum is simply a sum of blackbodies whose temperatures
are multiplied by Doppler factors; therefore, the signal from a bright region
of arbitrary temperature can be obtained by rescaling from a calculation 
with $kT=1$~keV. For each phase, the resulting spectra are folded through
a fiducial PCA response matrix, which we generated for PCU 2 during 
December 1999 (gain epoch 3), in order to obtain predicted 
light curves that can be compared directly to the observations. We analyze
these light curves in the seven energy
\begin{figure*}[thb]
%\epsscale{1.0}
%\plotone{f4.eps}
\centerline{\epsfig{file=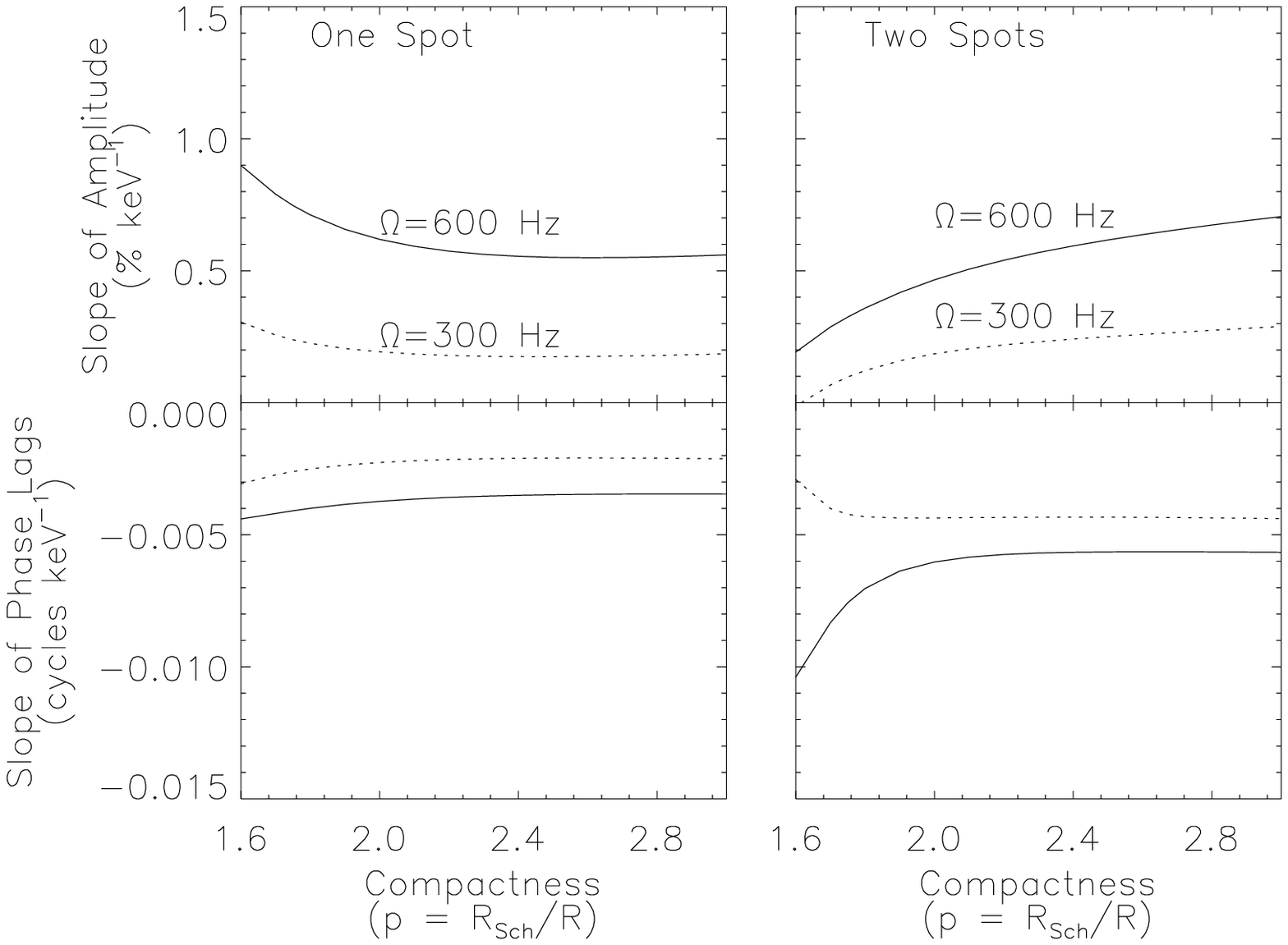,width=0.725\linewidth}}
\caption{Effects of the compactness $p=R/R_{\rm Sch}$ on the predicted
slopes of the amplitude and phase as a function of energy, for two
values of the spin $\Omega$, and considering both one and two bright 
regions. Here, $\alpha = \beta = 90^\circ$ and $\rho = 60^\circ$.}
\label{fig:compact}
\end{figure*}
 channels used in 
Section~\ref{sec:obs}, using the same Fourier techniques to measure the
amplitudes of the oscillations, and a least-squares fit to measure the 
phases. The amplitudes and phases we report are measured for the Fourier 
component at the spin frequency of the star when one hot spot is modeled,
and at twice the spin frequency when two antipodal spots are modeled.

\subsection{Energy Dependence of the Oscillations}

In Figure~\ref{fig:edep} we show the effects of the rotation of the neutron 
star on the energy-dependent profiles of the oscillations ({\it top panel}), 
and on the corresponding amplitudes ({\it middle panel}) and phases 
({\it bottom panel}) from a single circular bright region of temperature 
$kT_\infty = 2.3$~keV at infinity. Here, the bright region covers half the 
star ($\rho = 90^\circ$), is centered at the equator 
($\alpha = 90^\circ$), and is viewed along the equator ($\beta = 90^\circ$).
The star has a compactness $p = 2.5$ and spin $\Omega = 600$~Hz.

Stellar rotation introduces two main effects: the fractional amplitudes are 
larger in the higher energy bands, and the pulses at higher energies precede 
those at lower energies ({\it top panel} of Fig.~\ref{fig:edep}). 
Both effects occur because 
the spectrum is steepest at high energies ($\approx 20 $~keV), so that a 
slight Doppler shift in the apparent photon energy results in a significant 
change in the flux received. This not only yields a larger fractional amplitude
at $\approx 20$~keV, but also causes the hard pulse to peak as the spot rotates 
toward the observer, before the observer sees the largest solid angle from the 
spot at lower energies.

We quantify the energy dependence in Figure~\ref{fig:edep} in the same 
manner as in Section~2, by measuring the 
fractional \rms\ amplitudes ({\it middle panel}) and phases 
({\it bottom panel}) of the oscillations in each energy band. 
We perform a linear least-squares fit to the middle 5 energy channels 
(2--20~keV) from our simulations to determine the slopes 
of the amplitudes and phases as a function of energy. 
For the example in Figure~\ref{fig:edep}, the amplitude increases by 
approximately 0.4\% keV$^{-1}$, while the phases of the higher-energy photons 
lead those at lower-energy by about $0.003$ cycles keV$^{-1}$.

In Figure~\ref{fig:compact} we display how the amplitudes ({\it top panels}) 
and phases ({\it bottom panels}) of the
oscillations depend on the spin and compactness of the neutron star. 
The {left panels} illustrate the dependencies for the case of a single 
bright region, while the {right panels} illustrate the case of two 
antipodal regions. The bright regions have a constant temperature 
$kT_\infty = 2.3$~keV at infinity and a size $\rho = 60^\circ$, are
located at $\alpha = 90^\circ$, and are viewed along $\beta = 90^\circ$.
The rest of the star is assumed to be dark. 

We find that the energy dependence
of both the amplitudes and phases are considerably stronger when the 
neutron star is spinning rapidly ($\Omega = 600$~Hz; solid line), simply 
because the Doppler shifts are more pronounced. 
However, the dependence on compactness in Figure~\ref{fig:compact} is more 
complex. At larger compactness, $p>2.5$, the magnitudes 
of the amplitude increase are very similar for one or two bright regions,
although the phase lags are about 75\% larger from two spots. 
At small compactness ($p < 1.7$), the energy dependence of the pulse profiles
from one and two bright regions behave quite differently, because the bright 
region is strongly lensed on the opposite side of the star from the observer 
\citep{pfc83}. 
The lensing causes the relative phases at which 
({\it i}) the observer sees the largest solid angle from the emitting region 
and ({\it ii}) the Doppler effects are strongest to differ for the 
Fourier components at the spin frequency and its harmonic.
We use $p=2.5$ throughout the rest of our simulations, which corresponds
to a 1.4 \msun\ star with a 10~km radius.

\begin{inlinefigure}
%\epsscale{0.8}
%\plotone{f5.eps}
\centerline{\epsfig{file=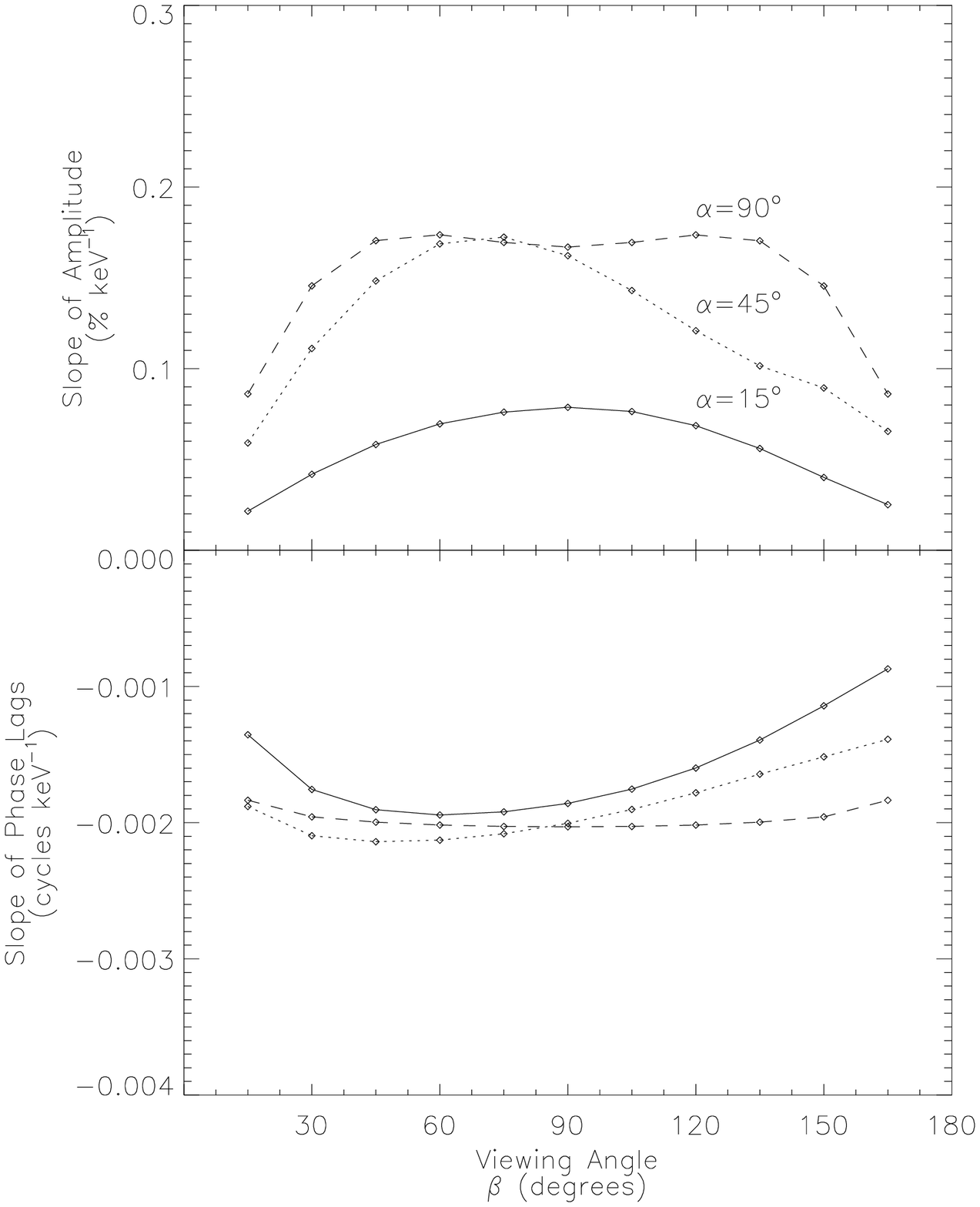,width=\linewidth}}
\caption{
Effects of changing the viewing angle $\beta$ ($x$-axis) and the
position of the bright region $\alpha$ (solid, dotted, and dashed lines) 
on the slopes of the amplitude and phase as a function of energy. A single
bright region with $\rho=60^\circ$ was modeled, on a $p=2.5$ neutron 
star with $\Omega = 300$~Hz. }
\label{fig:angles}
\end{inlinefigure}

We also find that as the size $\rho$ of the bright regions increases, the 
slope of the amplitude as a function of energy decreases monotonically. 
This is due to the facts
that ({\it i}) the amplitudes of the oscillations decrease in general, and 
({\it ii}) points on the bright regions have a range of velocities with 
respect to the observer, which damps the magnitude of Doppler effects.
The phase lags do not vary as a function of the size of the bright regions,
because the phase of the peak of the pulse is not changed by the 
effective background provided by the regions of a larger spot that are 
not Doppler-shifted.

In Figure~\ref{fig:angles}, we illustrate the effects of changing the
viewing angle ($\beta$) and position ($\alpha$) of one bright region on the 
energy dependence of the amplitudes and phases. Here we take
$\rho = 90^\circ$ and $kT_\infty = 2.3$~keV, and assume 
that the rest of the star is dark. The neutron star has $p=2.5$ and 
$\Omega = 300$~Hz. We find that the slopes of both the phases and amplitudes 
generally 
decrease in magnitude as the bright region or the observer are moved closer 
to the poles, because the relative velocity of the spot and the observer 
decreases. The slope of the fractional amplitude as a function of energy 
decreases more rapidly because the amplitude of the oscillation decreases,
just as occurs when the size of the bright region is increased.

As the temperature of the bright regions increases, the slopes of the 
phase lags and amplitudes flatten very slightly, because the peak of 
the energy spectrum moves to higher energies. However, if the rest of 
the neutron star 
\begin{inlinefigure}
%\epsscale{0.7}
%\plotone{f6.eps}
\centerline{\epsfig{file=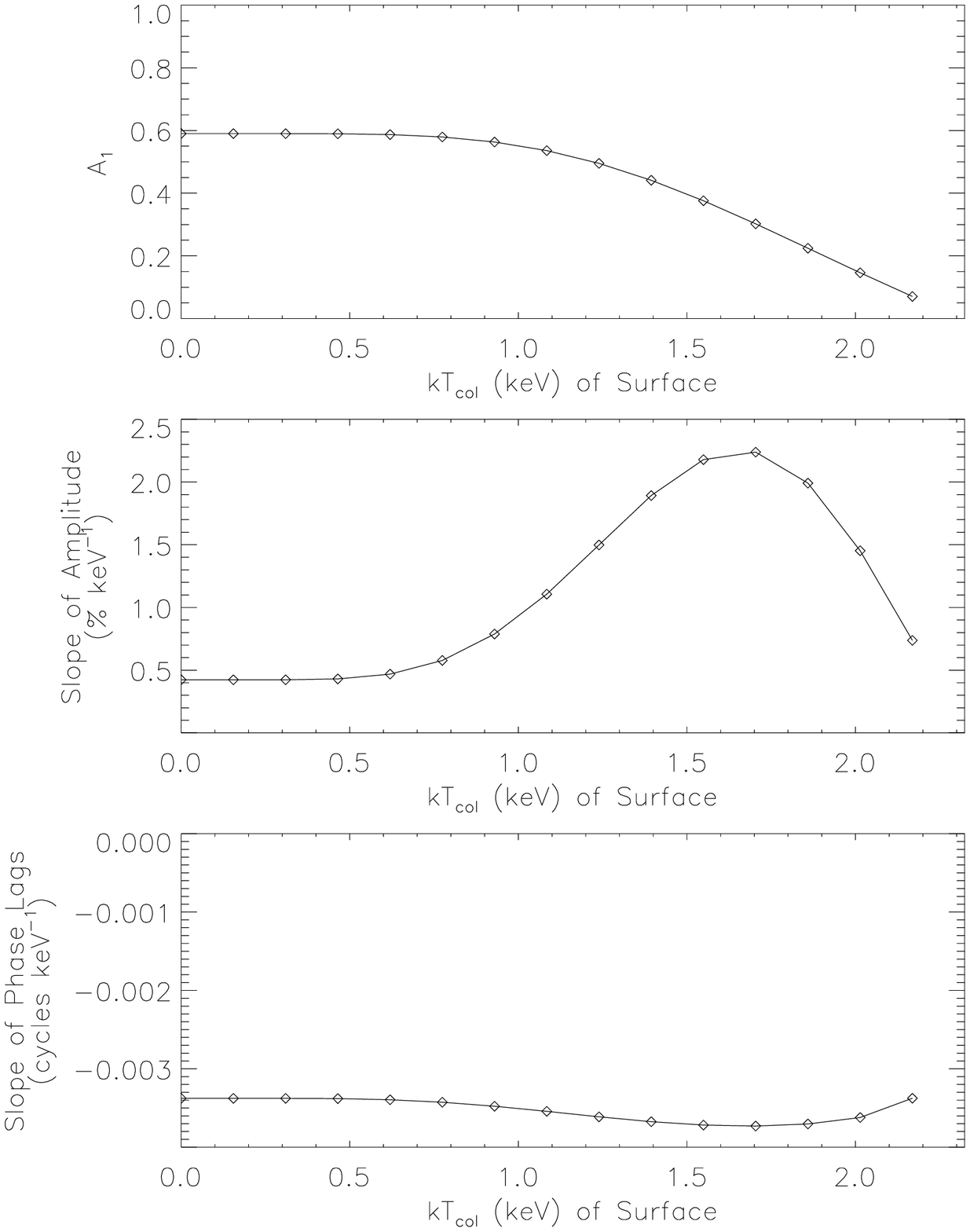,width=\linewidth}}
\caption{Effects of changing the temperature of the rest of the
neutron star, while keeping the bright region at a constant temperature
of $kT_\infty = 2.3$~keV as observed at infinity. We have taken 
$\rho=\alpha=\beta=90^\circ$, $\Omega=600$~Hz, and $p=2.5$. 
{\it Top panel}: The 
amplitude of the oscillation in the full PCA bandpass decreases 
monotonically as the temperature of the surface approaches that of the 
bright region. {\it Center panel}: The slope of the amplitude as a 
function of energy
reaches a maximum when the rest of the star emits within the PCA bandpass,
such that at low energies the oscillation is muted by a phase-independent 
background flux, while at high energies the emission is only observed 
from the hot region.
{\it Bottom panel}: Varying the temperature contrast only slightly affects the
slope of the phase as a function of energy.}
\label{fig:tcontrast}
\end{inlinefigure}

\noindent
also emits within the bandpass of the detector, the energy
dependence of the oscillations can become quite dramatic, as we illustrate
in Figure~\ref{fig:tcontrast}. The slope induced by the temperature contrast 
is more than five times larger than that produced by Doppler motion alone 
(i.e., when $kT_{\rm col} = 0$). At low energies, the cooler regions on 
the neutron star contribute a constant background flux to the oscillations, 
decreasing their fractional amplitude. At high energies, where the cool 
regions of the neutron star contribute little flux, the amplitude of an
oscillation is similar to that which one would see if the neutron star
were dark. The largest slope attainable increases as the size of the hot
region $\rho$ decreases. On the other hand, 
the phases of the oscillations are weakly
dependent on the temperature contrast from the neutron star, for the same 
reasons that the phases are not affected greatly by varying the sizes and 
positions of the bright regions. Thus, the phase lags provide the best 
measure of the apparent velocity of the bright region.

\newpage
\section{Discussion\label{sec:comp}}

\subsection{Amplitude as a Function of Energy}

In all of the burst oscillations, we find that the observed increase in 
amplitude as a function of photon energy is 
about 50\% larger than is expected from Doppler effects alone. In
the sources with $\approx 600$~Hz oscillations, \sixb\ and \mxbecl, the 
amplitude increases by up to 0.8--0.9~\%~keV$^{-1}$ (see 
Table~\ref{tab:ampve} and Fig.~\ref{fig:ampve}), while a single
bright region on a star spinning at 600~Hz can yield a  
0.6~\%~keV$^{-1}$ increase, and two antipodal regions on a star with 
$\Omega = 300$~Hz can give rise to a 0.3\% keV$^{-1}$ increase
(Fig.~\ref{fig:compact}).
Similarly, the amplitudes of the $\approx 300$~Hz oscillations from 
\sevenb\ and \slowb\ increase by 0.5--0.6\%~keV$^{-1}$, compared to the 
0.2\%~keV$^{-1}$ increase that might be expected from a bright region on a 
neutron star with $\Omega = 300$~Hz. These differences are significant, as
our uncertainties on the measured slopes are less than $0.01$\%~keV$^{-1}$.

It is therefore likely that the observed increase in amplitude as a function 
of energy occurs because the rest of the neutron star also emits in the 
the lower energy bands of the PCA (Fig.~\ref{fig:tcontrast}), thus
reducing the flux variations at these energies. To estimate the 
magnitude of the temperature contrast that could produce these oscillations, 
we use the constraints on the sizes 
and locations of bright regions that are consistent with the observed
oscillation profiles from \citet{moc02}. Using the lack of harmonic 
signals, \citet{moc02} show that the pulsations can only be produced by 
({\it i}) a single bright 
region that covers nearly half the neutron star ($\rho \approx 90^\circ$), 
({\it ii}) a single bright region that forms within $\alpha < 20^\circ$ 
from the rotational pole, 
or ({\it iii}) two antipodal bright regions that form within a few degrees 
of the rotational equator $\alpha \approx 90^\circ$. 

A single $\rho = 90^\circ$ bright region (case {\it i}) would produce
oscillations with an \rms\ amplitude of 55\% if the rest of 
the star were dark (Fig.~\ref{fig:tcontrast}, {\it top panel}). Since the 
observed amplitudes of the oscillations are 3--5 times smaller
even in the highest energy band, it is necessary to assume that the other 
half of the neutron star also emits, but at a lower temperature, in order 
to reproduce the observed 
oscillations. A temperature contrast of about 0.15~keV between the two 
hemispheres would produce oscillations with an amplitude of 5--10\% in 
the full PCA bandpass (Table~1 in Muno et al.\ 2002b) and an amplitude 
increase of 0.7\% keV$^{-1}$ (Table~\ref{tab:ampve}). In cases ({\it ii})
and ({\it iii}), on the other hand, the sizes of the spots are not 
well-constrained, and thus one
can reproduce the observed amplitudes with either small hot spots and
a small temperature contrast, or with large bright regions and a larger
temperature contrast (compare Fig.~\ref{fig:tcontrast} with Fig.~4 
in Muno et al.~2002b).

Finally, we note that the increases in the amplitudes of the burst oscillations
as a function of energy are not a generic feature of all coherent pulsations
from LMXBs. The amplitudes of the persistent pulsations from the 
accretion-powered millisecond X-ray pulsars \saxms\ and \asmms\ actually
decrease in amplitude between 2--10~keV. This suggests that a different 
emission mechanism or geometry may play a role in the generation of 
pulsations in these two classes of LMXBs.

\subsection{Phase as a Function of Energy}

As can be seen from Table~\ref{tab:phave} 
and Figure~\ref{fig:phave}, the phase lags observed from all of the combined 
pulse profiles are consistent either with being 
constant as a function of energy, or with increasing slightly such that the 
hard pulse arrives after the soft pulse. Only one set of oscillations, 
observed from \sixb\ in gain epoch 5, is consistent with a negative slope,
and in this case an energy-independent phase is also an acceptable fit.
The magnitudes of the phase lags are on order 0.5--0.1 cycles between 
2--20~keV, which translates to a time delay of $\approx 150$~$\mu$s for
spin frequencies between 300--600~Hz.
Thus, the signs of the observed phase lags are generally opposite the
negative phase lags that are expected from Doppler effects, although the
magnitudes of the lags are comparable. 
In contrast, the sign of the phase lags observed
from the accretion-powered millisecond X-ray pulsars \saxms\ and \asmms\ 
are consistent with those that would be expected from Doppler effects, 
as the pulse at 10~keV is observed to precede that at 2~keV by 
200--700~$\mu$s \citep{cmt98,gal02}. 

To understand the sign of the lags, we consider whether interactions 
between photons from the neutron star and material ({\it i}) in the 
accretion disk or ({\it ii}) in a hot corona could modify the energy 
dependence of the phases of
the oscillations. In the first case, photons that get scattered
by the accretion disk can acquire a Doppler shift from the motion of 
the orbiting material. However, since the disk is most likely to rotate in 
the same direction as the neutron star, the soft pulse would lag behind the 
hard \citep{ss01}, which is still inconsistent with the data.
In the second case, a corona of hot electrons could induce hard phase 
lags if they inverse-Compton scatter photons from the 
neutron star to higher energies \citep[e.g.,][]{mil95}. Indeed, the hard 
X-rays ($> 10$~keV) observed between bursts from neutron star LMXBs are 
thought to originate from such a corona \citep[e.g.,][]{bar00}. 
It is not clear whether such a corona also exists during X-ray bursts, 
%much less whether its temperature and 
%optical depth could produce the hard phase lags that we observe. 
%The hard pulse will lag behind the soft one only if the soft photons are 
%observed almost directly, while the hard photons scatter off of electrons 
%many more times before emerging at high energies \citep{mil95}. 
%Determining whether this occurs 
so evidence for it must be sought via simultaneous modeling of the 
burst spectra and the oscillation profiles. We also note that a 
hot corona of electrons could reduce the harmonic content of the
oscillation profiles \citep[e.g.,][]{mil00,moc02}.
%and in turn loosen the constraints on the geometry of the oscillations 
%determined in \citet{moc02}.

\section{Conclusions}

We have examined the energy-dependence in the amplitudes and phases of a 
sample of 51 burst oscillations using data from the \rxte\ PCA 
(Table~\ref{tab:sum}). We found that the 
fractional \rms\ amplitudes of the oscillations increase as a function of 
energy by 0.25\%~keV$^{-1}$ to 0.9\%~keV$^{-1}$ between 3--20~keV 
(Fig.~\ref{fig:ampve}
and Table~\ref{tab:ampve}). We then modeled the oscillations
as flux variations arising from temperature patterns on the surfaces of
rapidly rotating neutron stars, and calculated the Doppler effects on
the energy-resolved light curves. Comparing the models with the data, 
we found that the observed slope of the energy dependence is generally
larger than would be expected from Doppler effects 
alone (Fig.~\ref{fig:edep}). However, it can be reproduced by assuming 
that cooler regions of the neutron star also emit in the low energy bands
of the PCA (Fig.~\ref{fig:tcontrast}). 

We also found that the high-energy pulses generally lag behind the 
low-energy pulses in the observed light curves by 0.002(1) cycles rad$^{-1}$ 
to 0.007(3) cycles rad$^{-1}$ between 3--20~keV (Fig.~\ref{fig:phave} and 
Table~\ref{tab:phave}). In contrast, Doppler effects should 
cause the high-energy pulses to precede the soft pulses by 
0.003 cycles~keV$^{-1}$  (Fig.~\ref{fig:edep}). Hard lags like those 
observed could conceivably be produced if the signals from the 
surfaces of these neutron stars are re-processed by scattering in a
hot corona of electrons. However, it is not clear whether such a corona
exists during the declines of the thermonuclear bursts that 
produce these oscillations. Further observational and theoretical study 
is warranted to address this issue.

\acknowledgments{We thank Dimitrios Psaltis for helpful discussions, and 
Duncan Galloway, Pavlin Savov, and Derek Fox for important contributions 
to the data analysis underlying this work. 
M.M. and D.C. were supported in part by NASA, under contract 
NAS 5-30612 and grant NAG 5-9184.
F.\"O. acknowledges support by NASA through Hubble Fellowship grant
HF-01156 from the Space Telescope Science Institute, which is operated
by the Association of Universities for Research in Astronomy, Inc.,
under NASA contract NAS 5-26555.}

\begin{deluxetable}{cccc}
\tabletypesize{\scriptsize}
\tablecolumns{4}
\tablewidth{0pc}
\tablecaption{\rxte/PCA Channel-to-Energy Conversion\label{tab:ebounds}}
\tablehead{
\colhead{} & \multicolumn{3}{c}{Energy (keV)} \\
\colhead{} & \multicolumn{3}{c}{during Epoch} \\
\colhead{Channel} & \colhead{3} & \colhead{4} & \colhead{5} \\
}
\startdata
5  & 2.3 & 2.5 & 2.5 \\ 
13 & 5.1 & 5.9 & 5.7\\
18 & 6.9 & 8.0 & 7.8\\
28 & 10  & 12 & 12\\
36 & 13  & 16 & 15\\
53 & 20  & 23 & 22\\
\enddata
\tablecomments{Boundaries for the seven energy intervals used in this
paper, where the first interval is 0-5, and the last 53-255. 
See text for explanation of gain epochs.}
\end{deluxetable}

\begin{deluxetable}{lccccc}
\tabletypesize{\scriptsize}
\tablecolumns{6}
\tablewidth{0pc}
\tablecaption{Summed Profiles of Oscillations\label{tab:sum}}
\tablehead{
\colhead{Source} & \colhead{Frequency} & \colhead{Gain} & \colhead{No.} & 
\colhead{Total} & \colhead{Bkgd.} \\
\colhead{} & \colhead{(Hz)} &\colhead{Epoch} & \colhead{Osc.} & 
\colhead{Counts} & \colhead{Counts}
}
\startdata
\sixb\ & 581 & 3 & 6 & 312814 & 46871 \\
 & & 4 & 5 & 219356 & 23767 \\
 & & 5 & 4 & 317775 & 39453 \\[5pt]
\mxbecl\ & 567 & 4 & 3 & 27782 & 5761 \\[5pt]
\sevenb\ & 329 & 3 & 7 & 547076 & 47668 \\
 & & 5 & 1 & 38883 & 5236 \\[5pt]
\slowb\ & 363 & 3 & 12 & 552565 & 70491\\
 & & 4 & 6 & 291955 & 69510 \\[5pt]
\ksxrb\ & 321 & 3 & 4 & 249986 & 43025 \\[5pt]
\aqlxone\ & 549 & 3 & 2 & 120894 & 5754 \\
 & & 4 & 1 & 133784 & 6736 \\
\enddata
\end{deluxetable}

\begin{deluxetable}{lcccc}
\tabletypesize{\scriptsize}
\tablecolumns{5}
\tablewidth{0pc}
\tablecaption{Amplitudes of Burst Oscillations as a Function of Energy
\label{tab:ampve}}
\tablehead{
\colhead{} & \colhead{} & \colhead{Constant} &
\multicolumn{2}{c}{Linear Trend} \\
\colhead{Source} & \colhead{Gain} & \colhead{$\chi^2_\nu$} & 
\colhead{Slope} & \colhead{$\chi^2_\nu$} \\
\colhead{} & \colhead{Epoch} & \colhead{} & 
\colhead{[\% keV$^{-1}$]} & \colhead{} 
}
\startdata
\sixb\ & 3 & 17 & 0.448(6) & 0.9 \\
 & 4 & 16 & 0.390(6) & 1.1 \\
 & 5 & 108 & 0.889(5) & 5.4 \\[5pt]
\mxbecl\ & 4 & 5.9 & 0.777(18) & 0.7 \\[5pt]
\sevenb\ & 3 & 48 & 0.523(3) & 2.2 \\[5pt]
% & 5 & 112.5 & $> 0.999$ & 1.239(16) & 0.5 & 0.073\\
\slowb\ & 3 & 11 & 0.263(4) & 1.6 \\
 & 4 & 126 & 0.582(6) & 2.0 \\[5pt]
\ksxrb\ & 3 & 14 & 0.462(6) & 0.9 \\[5pt]
\aqlxone\ & 3 & 8.8 & 0.447(8) & 0.2 \\
% & 4 & 7.5 & 0.887 & 0.057(7) & 4.6 & 0.792\\
\enddata
\tablecomments{5 channels were used in computing $\chi^2_\nu$, so there were 
4 degrees of freedom assuming a constant amplitude, and 3 assuming a 
linear trend.}
\end{deluxetable}

\begin{deluxetable}{lcccc}
\tabletypesize{\scriptsize}
\tablecolumns{5}
\tablewidth{0pc}
\tablecaption{Phases of Burst Oscillations as a Function of Energy
\label{tab:phave}}
\tablehead{
\colhead{} & \colhead{} & 
\colhead{Constant} &
\multicolumn{2}{c}{Linear Trend}\\
\colhead{Source} & \colhead{Gain} & 
\colhead{$\chi^2_\nu$} & \colhead{Slope} & 
\colhead{$\chi^2_\nu$}  \\
\colhead{} & \colhead{Epoch} & 
\colhead{} & 
\colhead{[$10^{-3}$cycles keV$^{-1}$]} & \colhead{}
}
\startdata
\sixb\ & 3 & 1.0 & 1.92(14) & 0.6\\
 & 4 & 2.0 & 3.57(14) & 0.7 \\
 & 5 & 0.8 & $-$0.40(8) & 0.9\\[5pt]
\mxbecl\ & 4 & 2.6 & 7.74(25) & 0.3\\[5pt]
\sevenb\ & 3 & 2.4 & 1.97(10) & 1.6\\[5pt]
% & 5 & 1.0 & 0.086 & -0.0052(15) & 0.7 & 0.135\\
\slowb\ & 3 & 4.5 & 3.41(13) & 3.5\\
 & 4 & 3.5 & 4.31(11) & 0.8\\[5pt]
\ksxrb\ & 3 & 0.5 & 1.58(18) & 0.2\\[5pt]
\aqlxone\ & 3 & 0.8 & 3.04(29) & 0.5\\
% & 4 & 4.1 & 0.608 & 0.0213(19) & 2.7 & 0.566\\
\enddata
\tablecomments{5 channels were used in computing $\chi^2_\nu$, so there were 
4 degrees of freedom assuming a constant amplitude, and 3 assuming a 
linear trend.}
\end{deluxetable}


\begin{thebibliography}{0}
\bibitem[Alpar et al.(1982)]{alp82} Alpar, M. A., Cheng, A. F., Ruderman,
	M. A., \& Shaham, J. 1982, \nat, 300, 728
\bibitem[Barret \etal(2000)]{bar00} Barret, D., Olive, J. F., Boirin, L., 
	Done, C., Skinner, G. K., \& Grindlay, J. E. 2000, \apj, 
	533, 329
\bibitem[Bildsten(1995)]{bil95} Bildsten, L. 1995, \apj, 438, 852
\bibitem[Bildsten(1998)]{bil98} Bildsten, L. 1998, \apj, 501, L89
\bibitem[Braje et al.(2000)]{brr00} Braje, T. M., Romani, R. W., \&
	Rauch, K. P. 2000, \apj, 531, 447
\bibitem[Chandrasekhar(1960)]{cha60} Chandrasekhar, S. 1960, 
	{\it Radiative Transfer} (Dover)
%\bibitem[Cominsky \& Wood(1984)]{cw84} Cominsky, L. R. \& Wood, K. S. 1984,
%	\apj, 283, 765
\bibitem[Cui, Morgan, \& Titarchuk(1998)]{cmt98}  Cui, W., Morgan, E. H., 
	\& Titarchuk, L. G. 1998, 504, L27
%\bibitem[Cumming \& Bildsten(2000)]{cb00} Cumming, A. \& Bildsten, L. 2000, 
%	\apj, 544, 453
%\bibitem[Cumming et al.(2002)]{cum02} Cumming, A., Morsink, S. M., Bildsten,
%	L., Friedman, J. L., \& Holz, D. E. 2002, \apj, 564, 343
\bibitem[Ford(1999)]{for99} Ford, E. C. 1999, \apj, 519, L73
\bibitem[Fox(2000)]{fox00} Fox, D. W. 2001, PhD thesis, MIT
\bibitem[Fryxell \& Woosley(1982)]{fw82} Fryxell, B. A. \& Woosley, S. E.
	1982, \apj, 261, 332
%\bibitem[Galloway et al.(2001)]{gal01} Galloway, D. K., Chakrabarty, D.,
%	Muno, M. P., \& Savov, P. 2001, \apj, 549, L85
\bibitem[Galloway et al.(2002)]{gal02} Galloway, D. K., Chakrabarty, D., 
	Morgan, E. H., \& Remillard, R. A. 2002, \apj, 576, L137
\bibitem[Giles et al.(2002)]{gil02} Giles, A. B., Hill, K. M., Strohmayer, T. 
	E., \& Cummings, N. 2002, \apj, 568, 279
%\bibitem[Heyl(2002)]{hey02} Heyl, J. S. 2002, \mnras, submitted, 
%	astro-ph/0108450
\bibitem[Jahoda et al.(1996)]{jah96} Jahoda, K., Swank, J. H., Giles, A. B., 
	Stark, M. J., Strohmayer, T., Zhang, W., \& Morgan, E. H. 1996, 
	SPIE, 2808, 59
%\bibitem[Kuulkers et al.(2002)]{kul02} Kuulkers, E., Homan, J., van der Klis,
%	M., Lewin, W. H. G., \& M\'{e}ndez, M. 2002, \aap, 382, 947
\bibitem[Leahy \etal(1983)]{lea83} Leahy, D. A., Darbro, W., Elsner, R. F.,
        Weisskopf, M. C., Sutherland, P. G., Kahn, S., \& Grindlay, J. E.
        1983, \apj, 266, 160
%\bibitem[Lewin et al.(1993)]{lvt93} Lewin, W. H. G., van 
%	Paradijs, J., \& Taam, R. E. 1993, Space Sci. Rev., 62, 223
%\bibitem[London et al.(1984)]{lth84} London, R. A., Taam, R. E.,
%	\& Howard, W. M. 1984, \apj, 287, L27
\bibitem[Madej(1991)]{mad91} Madej, J. 1991, \apj, 376, 161
\bibitem[Manchester \& Taylor(1977)]{mt77} Manchester, R. N., \& Taylor,
    J. H. 1977, Pulsars(San Francisco: W. H. Freeman and Co.)
\bibitem[Miller(1995)]{mil95} Miller, M. C. 1995, \apj, 441, 770
\bibitem[Miller(1999)]{mil99} Miller, M. C. 1999, \apj, 515, L77
\bibitem[Miller(2000)]{mil00} Miller, M. C. 2000, \apj, 537, 342
\bibitem[Miller \& Lamb(1998)]{ml98} Miller, M. C. \& Lamb, F. K. 1998, 
	\apj, 499, L37
\bibitem[Miller et al.(1998)]{mlp98} Miller, M. C., Lamb, F. K.,
	\& Psaltis, D. 1998, \apj, 508, 791
%\bibitem[Muno et al.(2001)]{mun01} Muno, M. P., Chakrabarty, D., Galloway,
%	D. K., \& Savov, P. 2001, \apj, 553, L157
%\bibitem[Muno et al.(2000)]{mun00} Muno, M. P., Fox, D. W., Morgan, E. H., 
%	\& Bildsten, L. 2000, \apj, 542, 1016
\bibitem[Muno et al.(2002a)]{mun02} Muno, M. P., Galloway, D. K. Chakrabarty,
	D., \& Psaltis, D. 2002a, \apj, 580, 1048
\bibitem[Muno \etal(2002b)]{moc02} Muno, M. P., \"{O}zel, F., \& Chakrabarty, 
	D. 2002b, \apj, 581, 550
\bibitem[Nath et al.(2002)]{nss02} Nath, N. R., Strohmayer, T. E., \& 
	Swank, J. H. 2002, \apj, 564, 353
%\bibitem[\"{O}zel(2002)]{oze02} \"{O}zel, F. 2002, \apj, in press, 
%	astro-ph/021158
%\bibitem[\"{O}zel, Psaltis, \& Kaspi(2001)]{opk01} \"{O}zel, F., 
%	Psaltis, D., \& Kaspi, V. M. 2001, \apj\, 563, 255
\bibitem[Page(1995)]{page95} Page, D. 1995, \apj, 442, 273
\bibitem[Pechenick et al.(1983)]{pfc83} Pechenick, K. R., Ftaclas, C., \&
	Cohen, J. M. 1983, \apj, 274, 846
%\bibitem[Popham \& Sunyaev(2001)]{ps01} Popham, R. \& Sunyaev, R. 2001, 
%	\apj, 547, 355
\bibitem[Press \etal\ (1992)]{pre92} Press, W. H., Teukolsky, S. A.,
        Vetterling, W. T. \& Flannery, B. P. 1992, Numerical Recipes
        in C, 2nd Ed. (Cambridge: Cambridge University Press)
\bibitem[Radhakrishnan \& Srinivasan(1982)]{rs82} Radhakrishnan, V.
         \& Srinivasan, G. 1982, Curr. Sci., 51, 1096
\bibitem[Sazonov \& Sunyaev(2001)]{ss01} Sazonov, S. Y. \& Sunyaev, R. A. 
	2001, \aap, 373, 241
%\bibitem[Schoelkopf \& Kelley(1991)]{sk91} Schoelkopf, R. J. \& 
%	Kelley, R. L. 1991, \apj, 375, 696
%\bibitem[Spitkovsky et al.(2002)]{slu01} Spitkovsky, A., 
%	Levin, Y., \& Ushomirsky, G. 2002, \apj, 566, 1018
%\bibitem[Strohmayer(1999)]{str99} Strohmayer, T. E. 1999, \apj, 523, L51
\bibitem[Strohmayer(2001)]{str01} Strohmayer, T. E. 2001,
	Adv. Space. Res., 28, 511
\bibitem[Strohmayer et al.(1997)]{str97} Strohmayer, T. E., Jahoda, K., 
	Giles, A. B., \& Lee, U. 1997, \apj, 486, 355
\bibitem[Strohmayer \& Markwardt(1999)]{sm99} Strohmayer, T. E. \& 
	Markwardt, C. B. 1999, \apj, 516, L81
\bibitem[Strohmayer \& Markwardt(2002)]{sm02} Strohmayer, T. E. \&
	Markwardt, C. B. 2002, \apj, submitted
%\bibitem[Strohmayer et al.(1997b)]{szs97} Strohmayer, T. E., Zhang, 
%	W., \& Swank, J. H. 1997b, \apj, 487, L77
\bibitem[Strohmayer et al.(1998b)]{str98b} Strohmayer, T. E., Zhang, 
	W., Swank, J. H., \& Lapidus, I. 1998b, \apj, 503, L147
\bibitem[Strohmayer et al.(1998a)]{str98a} Strohmayer, T. E., Zhang, 
	W., Swank, J. H., White, N. E., \& Lapidus, I. 1998a, \apj, 498, L135
\bibitem[Strohmayer et al.(1996)]{str96} Strohmayer, T. E., Zhang, W., Swank, 
	J. H., Smale, A., Titarchuk, L., Day, C., \& Lee, U. 1996, \apj, 
	469, L9
%\bibitem[van der Klis(2000)]{vdk00} van der Klis, M. 2000, \araa, 38, 717
\bibitem[Vaughan et al.(1994)]{vau94} Vaughan, B. A. et al. 1994, \apj, 
	435, 362
\bibitem[Weinberg et al.(2001)]{wml01} Weinberg, N., Miller, M. C., \&
 	Lamb, D. Q. 2001, \apj, 546, 1098
%\bibitem[White \& Zhang(1997)]{wz97} White, N.~E. \& Zhang, W. 1997,
%	\apj, 490, L87
\end{thebibliography}
\end{document}